\documentclass[english,conference]{IEEEtran}
\usepackage[T1]{fontenc}
\usepackage[latin9]{inputenc}
\usepackage{color}
\usepackage{array}
\usepackage{multirow}
\usepackage{amsmath}
\usepackage{amssymb}
\usepackage{graphicx}

\makeatletter

\providecommand{\tabularnewline}{\\}

\usepackage{cite}
\usepackage{bm}
\usepackage{algorithmic}
\usepackage{algorithm}
\usepackage{graphicx}
\IEEEoverridecommandlockouts
\usepackage{balance}
\usepackage{epstopdf}
\usepackage{subfigure}
\usepackage[table]{xcolor}
\makeatother

\usepackage{babel}
\begin{document}
\title{Regional Resource Management for Service Provisioning in LEO Satellite
Networks: A Topology Feature-Based DRL Approach}
\author{\IEEEauthorblockN{Chenxi Bao$^{\dagger}$, Di Zhou$^{\dagger\ast}$,
Min Sheng$^{\dagger}$, Yan Shi$^{\dagger}$, Jiandong Li$^{\dagger}$,
and Zhili Sun$^{\ddagger}$ }\IEEEauthorblockA{\textsuperscript{}$^{\dagger}$State
Key Laboratory of Integrated Service Networks, Xidian University,
Xi'an, Shaanxi, 710071, China\\
$^{\ddagger}$Institute of Communication Systems (ICS), 5G/6G Innovation
Centre, School of Computer Science and Electronic\\
Engineering, Faculty of Engineering and Physical Sciences, University
of Surrey, GU27XH Guildford, UK\\
$^{*}$Email: zhoudi@xidian.edu.cn}\vspace{-2.2em}\thanks{This work was supported in part by National Natural Science Foundation
of China (Grant Nos. 62422114, 62461160329, 62371360, 62495020), Young
Elite Scientists Sponsorship Program by CAST (No.2022QNRC001), and
the Fundamental Research Funds for the Central Universities under
Grant QTZX25101.}}
\maketitle
\begin{abstract}
\textcolor{black}{Satellite networks with wide coverage are considered
natural extensions to terrestrial networks for their long-distance
end-to-end (E2E) service provisioning. However, the inherent topology
dynamics of low earth orbit satellite networks and the uncertain network
scales bring an inevitable requirement that resource chains for E2E
service provisioning must be efficiently re-planned.}\textcolor{blue}{{}
}Therefore, achieving highly adaptive resource management is of great
significance in practical deployment applications. \textcolor{black}{This
paper first designs a regional resource management (RRM) mode and
further formulates the RRM problem that can provide a unified decision
space independent of the network scale. Subsequently, leveraging the
RRM mode and deep reinforcement learning framework, we develop a topology
feature-based dynamic and adaptive resource management algorithm to
combat the varying network scales. The proposed algorithm successfully
takes into account the fixed output dimension of the neural network
and the changing resource chains for E2E service provisioning. The
matched design }of the service orientation information and phased
reward function effectively improves the service performance of the
algorithm under the RRM mode. \textcolor{black}{The numerical results
demonstrate that the proposed algorithm with the best convergence
performance and fastest convergence rate significantly improves service
performance for varying network scales, with gains over compared algorithms
of more than 2.7\%, 11.9\%, and 10.2\%, respectively.}
\end{abstract}

\begin{IEEEkeywords}
\textcolor{black}{Satellite networks, resource management, service
performance optimization, topology feature learning.}\vspace{-0.6em}
\end{IEEEkeywords}

\section{Introduction\vspace{-0.3em}}

Nowadays, the rapid development of the economy and society has brought
about earth-shaking changes in the lifestyles of people all over the
world. Various emerging applications, \textcolor{black}{such as augmented
reality, virtual reality, and Internet of Things (IoT), are needed
by all kinds of users in all regions \cite{wang2022dynamic}.} However,
mobile communication systems that still rely on terrestrial infrastructure
have never been able to achieve global ubiquitous connectivity and
seamless services \cite{Zhou2024Intelligent}. \textcolor{black}{To
this end, researchers of the sixth-generation (6G) mobile communication
system are working hard to achieve this grand goal and have driven
the deployment of satellite networks based on low earth orbit satellites
(LEOs), such as OneWeb, Starlink, and Telesat \cite{Luo2024LEO}.}
Satellite networks with wide coverage are considered irreplaceable
in the long-distance transmission of services compared with existing
terrestrial networks or even air networks. Therefore, it is very important
and highly favored by users for satellite networks for end-to-end
(E2E) service provisioning.

In this case, users will access the LEO through a direct connection
mode that is different from the traditional satellite network that
provides information exchange between ground stations, and E2E service
provisioning will be achieved using the satellite network as the service-bearing
entity between any two nodes, such as user-to-user, user-to-server,
etc \cite{Voelsen2021}. However, due to the inherent orbital deployment
and high-speed movement of LEOs, inter-satellite links (ISLs) are
intermittently connected, which means that the network topology is
highly dynamic to inevitably re-plan the resource chains for E2E service
provisioning. Furthermore, the scale of satellite networks is highly
differentiated, ranging from tens to tens of thousands of LEOs, which
means that resource management algorithms should be highly adaptable
in practical deployment applications. 

Resource management for E2E service provisioning in satellite networks
has attracted more attention recently \cite{wang2023joint,Dong2022Intelligent,liu2021routing,tsai2022multi}.
The resource evolution relationship based on the satellite network
topology was modeled as a time-expanded graph, and the service flow
optimization problem on this graph was solved by leveraging the proposed
iterative heuristic algorithms in \cite{wang2023joint}, which is
a service provisioning scheme that directly plans E2E transmission
resource chains. To combat the dynamic network environment, \cite{Dong2022Intelligent}
applied deep reinforcement learning (DRL) to service provisioning
to optimize resource management strategy by selecting candidate resource
chains to meet resource constraints. However, the path-level decision
models in \cite{wang2023joint} and \cite{Dong2022Intelligent} are
difficult to cope with the varying network scale, and in light of
the fixed characteristics of the neural network output dimension,
the candidate resource chains are calculated in advance and the number
cannot be changed, which hinders the practical application value of
the algorithms. In \cite{liu2021routing}, the E2E service provisioning
was formulated as a series of next-hop selection processes, and a
resource chain planning algorithm based on network topology feature
extraction was proposed. Similarly, \cite{tsai2022multi} proposed
a DRL-based E2E service provisioning algorithm that can simultaneously
make next-hop decisions for multiple services. However, the input
of a global network structure in \cite{liu2021routing} and the actions
of fixed dimension related to the number of service requests in \cite{tsai2022multi}
are bound to lead to retraining under varying network environments.

This paper investigates the resource management for E2E service provisioning
in satellite networks to alleviate the limitations of network environment
uncertainty on the practical deployment and application of the algorithm.
There are two challenges as follows:
\begin{enumerate}
\item \textcolor{black}{How to formulate a resource management problem to
ensure the unification of decision spaces under varying network scales?}
\item \textcolor{black}{How to design a resource management algorithm to
alleviate the impact of dynamic network environments in its adaptability?}
\end{enumerate}
\setlength{\parindent}{1em}

To solve the above challenges, we propose a topology feature-based
dynamic and adaptive resource management (TF-DARM) algorithm to ensure
E2E service provisioning performance in different satellite networks.
The contribution of our work can be summarized as follows:
\begin{enumerate}
\item \textbf{Regional resource management mode}: We start from the orbital
deployment and movement of LEOs and clarify the similar topology features
of different satellite network environments. \textcolor{black}{Based
on this feature, we design a regional resource management (RRM) mode
that is different from conventional multi-resource chain E2E service
provisioning and further formulate the RRM problem that can provide
a unified decision space in different network scales.}
\item \textbf{TF-DARM algorithm}: \textcolor{black}{To combat the dynamic
network environments and avoid retraining, we develop the TF-DARM
algorithm to obtain a trained model independent of the network scale.
Specifically, the algorithm adopts a generalized action space to take
into account the fixed output dimensions of the neural network and
the changing resource chains for E2E service provisioning in various
network environments.}\textcolor{blue}{{} }Furthermore, the state with
service orientation information and the phased reward function are
designed to progressively guide service requests to approach the destination
node.
\end{enumerate}
\setlength{\parindent}{1em}

The remainder of this paper is organized as follows. Section \ref{sec:System-Model}
presents the system model. The RRM problem is formulated in Section
\ref{sec:Formulation}. In Section \ref{sec:TF-DARM-Algorithm-Design},
we convert the problem into a Markov Decision Process (MDP) and propose
the TF-DARM algorithm to solve it. The numerical results are shown
in Section \ref{sec:Simulations}, and finally, we conclude the paper
in Section \ref{sec:Conclusion}.\vspace{-0.2em}

\section{System Model\label{sec:System-Model}\vspace{-0.2em}}

In this section, we first describe the network model, and then introduce
the service request model, and delay model, respectively.\vspace{-0.3em}

\subsection{Network Model\label{subsec:Network-Model}\vspace{-0.3em}}

\begin{figure}
\begin{centering}
\includegraphics[scale=0.38]{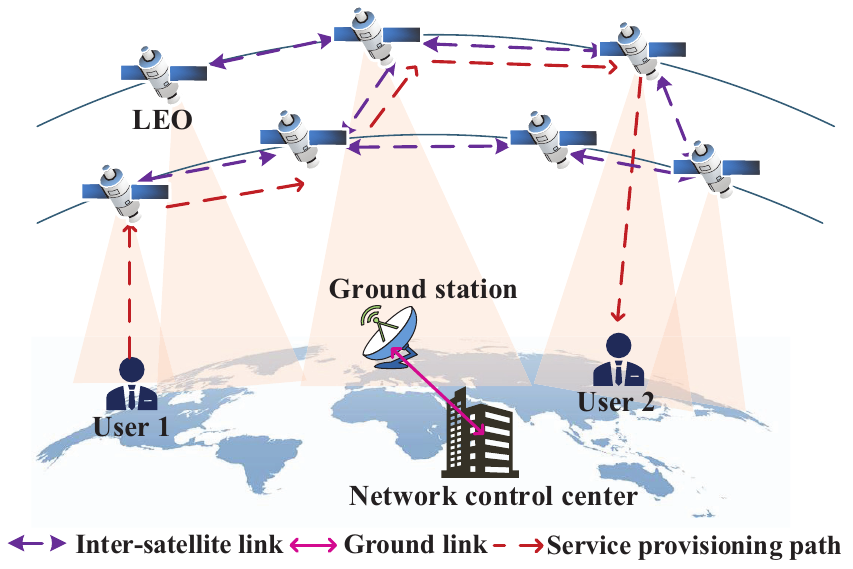}\vspace{-1.0em}
\par\end{centering}
\caption{\label{fig:Illustration-of-a}Illustration of a satellite network
scenario and service request transmitting.}
\vspace{-1.6em}
\end{figure}

We consider a typical satellite network scenario for E2E service provisioning,
which mainly includes a set of LEOs $\mathcal{L}=\left\{ i\left|i=1,2,\cdots,\left|\mathcal{L}\right|\right.\right\} $
and one network control center (NCC), as shown in Figure \ref{fig:Illustration-of-a}.\textcolor{red}{}\textcolor{black}{{}
Furthermore, this paper considers the satellite network realizes service
bearing between the source and the destination nodes of service requests.
For example, as shown in Figure \ref{fig:Illustration-of-a}, ``User
1'' transmits its services to ``User 2'' through the satellite
network. It should be noted that the ground stations summarize resources
and service request information from LEOs to the NCC for training
resource management models, and we }assume that the effectiveness
of network resources and service requirement information can be guaranteed
to study the impact of resource management strategies on the service
performance of satellite networks.\textcolor{black}{{} }

Due to the orbiting movements, the connection relationship between
two LEOs is time-varying. We can observe that the topology of satellite
networks directly affects resource chains for E2E service provisioning.
Meanwhile, we also observe that satellite networks are different from
terrestrial and air networks. The characteristics of their orbital
deployment make the \textquotedbl one-satellite four-links\textquotedbl{}
mode widely used in the establishment of ISLs \cite{Song2024Topological}.
\textcolor{black}{Specifically, in light of the laser ISL has been
widely used in constellations, such as the Starlink constellation,
this paper considers LEOs with laser communication devices to exchange
with each other \cite{zhu2022laser}. According to the ISL establishment
rule, each LEO is equipped with four laser terminals, which are used
to establish two intra-orbit plane laser ISLs and two inter-orbit
plane laser ISLs, respectively \cite{Wang2024Dynamic}.}\textcolor{blue}{{}
}The set of directional ISLs is denoted by $\mathcal{ISL}=\left\{ \left(i,j\right)\left|i,j\in\mathcal{L}\right.\right\} $.
\textcolor{black}{Furthermore, we divide the planning cycle into a
set of time slots, denoted by $T$, and the interval of each time
slot $t\in T$ is fixed as $\tau$.}\vspace{-0.6em}
\begin{figure*}
\begin{centering}
\includegraphics[scale=0.39]{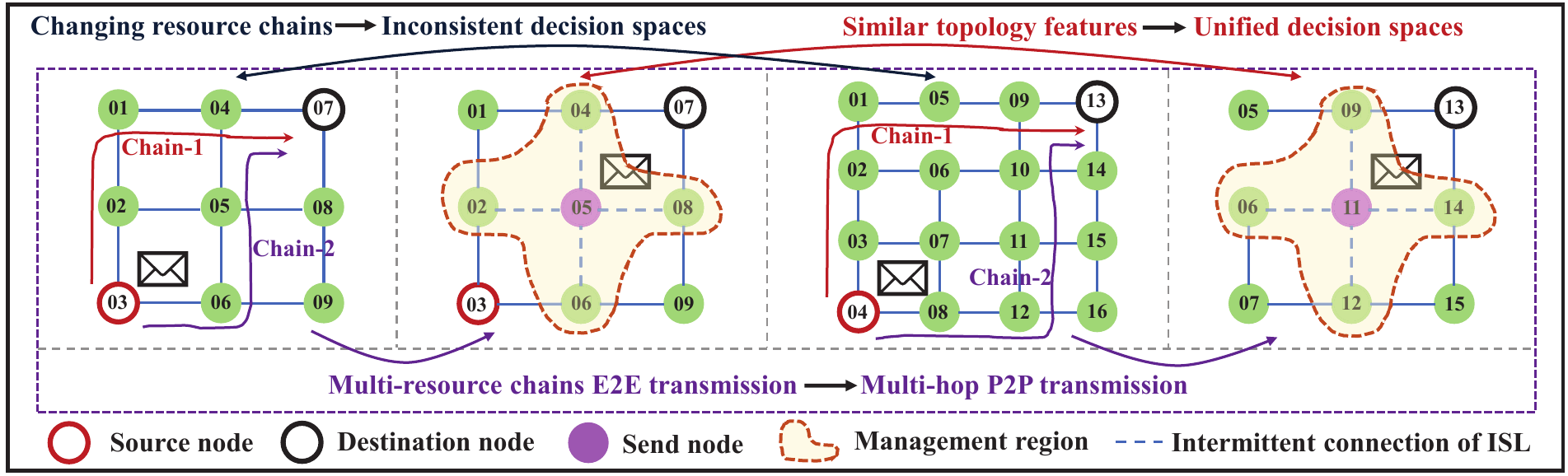}\vspace{-1.0em}
\par\end{centering}
\caption{\label{fig:Illustration-of-resource}Illustration of resource management
mode for E2E service provisioning in satellite networks.}
\vspace{-1.6em}
\end{figure*}

\subsection{Service Request Model\label{subsec:Service-Reuqest-Model}\vspace{-0.3em}}

\textcolor{black}{This paper focuses on the E2E service provisioning
process after service requests arrive at LEOs, considers the batch
service provisioning of service requests, and sets its batch service
period to a time slot \cite{Wang2020SFC}. Furthermore, we assume
that a batch of service requests with different service deadlines
arrives }in each service period, and we determine the service deadline
of each service request based on its arrival time and delay requirement\footnote{This paper considers the causality of on-board storage, i.e., service
requests in the current time slot arriving are served in the next
time slot.}. In addition, the service process of service requests may last for
multiple time slots, and service requests across time slots will be
combined with newly arrived service requests into one batch to be
served.

Furthermore, we construct a service request sequence, denoted by $\mathcal{Q}=\left\{ q\left|q=1,2,\ldots\left|\mathcal{Q}\right|\right.\right\} $
and define each service request $q$ denoted as $q=\left(S_{q},D_{q},At_{q},B_{q}^{\textrm{req}},L_{q}^{\textrm{req}}\right),$
where $S_{q}$ and $D_{q}$ represent the source and destination node,
respectively. $At_{q}$ is the arrival time. $B_{q}^{\textrm{req}}$
and $L_{q}^{\textrm{req}}$ represent the data transmission requirement
and delay requirement, respectively.\vspace{-0.6em}

\subsection{Delay Model\label{subsec:Delay-Model}\vspace{-0.3em}}

\textcolor{black}{Considering the advantages of satellite networks
in long-distance transmission, service requests using satellite network-providing
service may go through multiple hops from the source node to the destination
node.}\textcolor{red}{{} }\textcolor{black}{The delay of the single-hop
consists of four parts, expressed as:}\textcolor{blue}{\vspace{-1.0em}}

\textcolor{black}{
\begin{equation}
L_{q}^{i}\left(t\right)\!=\!L_{\textrm{tran}}^{\left(i,j\right)}\left(q,t\right)\!+\!L_{\textrm{prop}}^{\left(i,j\right)}\left(q,t\right)\!+\!L_{\textrm{proc}}^{i}\left(q,t\right)\!+\!L_{\textrm{que}}^{i}\left(q,t\right).\vspace{-0.5em}
\end{equation}
Wherein, $L_{\textrm{tran}}^{\left(i,j\right)}\left(q,t\right)$ represents
the transmission delay, calculated as:}\vspace{-0.5em}\textcolor{black}{
\begin{equation}
L_{\textrm{tran}}^{\left(i,j\right)}\left(q,t\right)=\frac{B_{q}^{\textrm{req}}}{\mathcal{R}^{\left(i,j\right)}\left(t\right)},\vspace{-0.5em}
\end{equation}
where $\mathcal{R}^{\left(i,j\right)}\left(t\right)$ indicates the
achievable transmission rate (in bps) of ISL from LEO $i$ to LEO
$j$ in the $t$-th time slot. $L_{\textrm{prop}}^{\left(i,j\right)}\left(q,t\right)$
represents the propagation delay, expressed as:}\textcolor{blue}{\vspace{-0.5em}}\textcolor{black}{
\begin{equation}
L_{\textrm{prop}}^{\left(i,j\right)}\left(q,t\right)=\frac{d^{\left(i,j\right)}\left(t\right)}{\mathcal{C}},\vspace{-0.5em}
\end{equation}
where $d^{\left(i,j\right)}\left(t\right)$ is the propagation distance
(in km), and $\mathcal{C}$ is the speed of light (in km/s). $L_{\textrm{proc}}^{i}\left(q,t\right)$
represents the processing delay, and we consider that each service
request is compressed when arriving at the LEO to reduce the requirement
for the link rate. $L_{\textrm{que}}^{i}\left(q,t\right)$ represents
the processing delay and the queuing delay.}\vspace{-0.3em}

\section{Formulation\label{sec:Formulation}\vspace{-0.3em}}

In this section, we first introduce the RRM mode. Then, we present
the resource, link, and service provision constraints based on RRM
mode. Finally, we formulate the proposed RRM problem based on these
constraints.\vspace{-0.3em}

\subsection{Regional Resource Management Mode\label{subsec:Regional-Resource-Management}\vspace{-0.3em}}

As we all know, trained neural network models can perform stably with
excellent performance in similar training environments. \textcolor{black}{However,
we also understand that varying network scales have dealt a severe
blow to the practicality of DRL-based resource management algorithms.}
In the resource management problem of satellite networks for E2E service
provisioning, the essence of its inability to cope with uncertainty
is the change of candidate resource chains, as shown in Figure \ref{fig:Illustration-of-resource}.
To this end, we design the RRM mode to ensure the unification of \textcolor{black}{decision
spaces} leveraging the local topology feature of satellite networks. 

Specifically, as shown in Figure \ref{fig:Illustration-of-resource},
in this mode, conventional multi-resource chains E2E transmission
decisions are converted into multi-hop point-to-point (P2P) transmission
decisions to complete the service provisioning. We divide the one-hop
reachable range centered on the send node into a region and determine
th\textcolor{black}{e decision space for }P2P transmission based on
the region. In light of the limited number of transceivers and the
widely adopted ISL establishment rules as described in Section \ref{subsec:Network-Model},
the local topology features of different satellite networks are similar,
which ensures that the maximum number of next-hop nodes that can be
selected at any send node is consistent. Besides, we also clarify
that due to the intermittent connection of ISL, the number of optional
next-hop nodes contained in the region changes dynamically. Based
on the above analysis, we can conclude that applying the RRM mode
to formulate the resource management problem can provide a unified
\textcolor{black}{decision space fo}r the solution algorithm based
on the DRL framework.\vspace{-0.3em}

\subsection{\textcolor{black}{RRM-Based Mode Constraints}\label{subsec:Based-RRM-Mode-Related}\vspace{-0.3em}}

\subsubsection{Communication Resource Constraint}

\textcolor{black}{The total data volume of transmitted service requests
cannot exceed the maximum data volume that can be carried by} \textcolor{black}{ISL
$\left(i,j\right)$, expressed as:}\vspace{-0.5em}\textcolor{black}{
\begin{equation}
\underset{q\in\mathcal{Q}}{\sum}B_{q}^{\textrm{req}}\cdot\!\xi_{q}^{\left(i,j\right)}\!\left(t\right)\!\leq\mathcal{R}^{\left(i,j\right)}\left(t\right)\!\cdot\!\tau,\forall\!\left(i,j\right),t,\vspace{-0.5em}\label{eq:com}
\end{equation}
}where $\xi_{q}^{\left(i,j\right)}\left(t\right)$ is a binary variable
for indicating whether the service request $q$ is transmitted on
the ISL $\left(i,j\right)$ in the $t$-th time slot, 1 if $q$ is
transmitted on the ISL $\left(i,j\right)$, 0 otherwise.

\subsubsection{Link Selection Constraint}

For any one RRM decision-making, this paper does not consider the
splitting of service requests, thus, only one ISL can be selected
for transmitting service requests, expressed as:\vspace{-0.5em}\textcolor{black}{
\begin{equation}
\underset{\left(i,j\right)\in\mathcal{ISL}}{\sum}\xi_{q}^{\left(i,j\right)}\left(t\right)\leq1,\forall\left(i,j\right),t.\vspace{-0.5em}
\end{equation}
}\vspace{-0.3em}
\begin{figure*}
\begin{centering}
\includegraphics[scale=0.43]{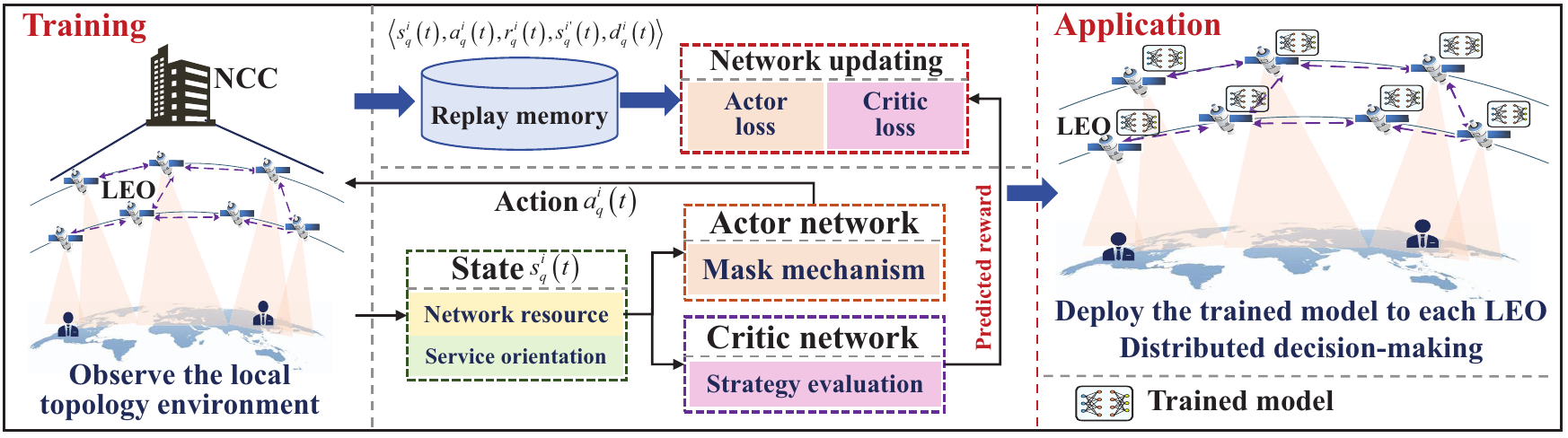}\vspace{-1.0em}
\par\end{centering}
\caption{\label{fig:An-overview-of}An overview of training and application
of the proposed TF-DARM algorithm.}
\vspace{-1.4em}
\end{figure*}

\subsubsection{Service Provisioning Constraints}

\textcolor{black}{The E2E service provisioning of any service request
that is successfully completed should satisfy: 1) the start node and
the end node are the source node and the destination node, respectively,
expressed as follows: }\vspace{-0.3em}
\begin{equation}
\underset{t\in T}{\sum}\underset{\left(S_{q},j\right)\in\mathcal{ISL}}{\sum}\!\xi_{q}^{\left(S_{q},j\right)\!}\left(t\right)=1,\!\forall\,q,\vspace{-0.6em}
\end{equation}
\vspace{-0.3em}
\begin{equation}
\underset{t\in T}{\sum}\underset{\left(i,D_{q}\right)\in\mathcal{ISL}}{\sum}\!\xi_{q}^{\left(i,D_{q}\right)}\!\left(t\right)=1,\!\forall\,q.\vspace{-0.6em}
\end{equation}
\textcolor{black}{and 2) the total time from the source to the destination
node cannot exceed the delay requirement, expressed as:}\vspace{-1.0em}\textcolor{black}{}

\begin{equation}
\begin{array}{cc}
\underset{t\in T}{\sum}\!\left(\underset{\left(i,j\right)\in\mathcal{ISL}}{\sum}\!\!\left(L_{\textrm{tran}}^{\left(i,j\right)}\!\left(q,t\right)\!+\!L_{\textrm{prop}}^{\left(i,j\right)}\!\left(q,t\right)\right)\cdot\!\xi_{q}^{\left(i,j\right)}\!\left(t\right)\!\!\!\!\!\!\!\!\!\!\!\!\!\!\!\!\!\!\!\!\!\!\!\!\!\!\!\!\!\!\!\!\!\!\!\!\!\!\!\!\!\!\right. & \vspace{-0.5em}\\
 & \left.\!\!\!\!\!\!\!\!\!\!\!\!\!\!\!\!\!\!\!\!\!\!\!\!\!\!\!\!\!\!\!\!\!\!\!\!\!\!\!\!\!\!\!\!\!\!\!\!\!\!\!\!\!\!\!\!\!\!+\!\underset{i\in\mathcal{L}}{\sum}\!\left(L_{\textrm{proc}}^{i}\!\left(q,t\right)\!+\!L_{\textrm{que}}^{i}\left(q,t\right)\right)\!\right)\!\leq\!L_{q}^{\textrm{req}},\forall\,q.
\end{array}\vspace{-0.5em}\label{eq:delay-1}
\end{equation}
Otherwise, it will be regarded as a service failure and will be deleted
from the on-board storage.\textcolor{black}{}

\textcolor{black}{Furthermore, we define }a binary variable\textcolor{black}{{}
$\mathcal{N}_{q}$} to indicate\textcolor{black}{{} whether the service
request $q$ is successfully served}, 1 if \textcolor{black}{successfully
served}, 0 otherwise.\vspace{-0.6em}

\subsection{Problem Formulation\label{subsec:Problem-Formulation}\vspace{-0.5em}}

\textcolor{black}{This work aims to ensure E2E service provisioning
performance in satellite networks by optimizing RRM strategies, endeavoring
to maximize the number of successfully accomplished service requests
while satisfying constraints on the network resources and QoS requirement
of services. Mathematically, the RRM problem is expressed as:}\vspace{-0.5em}
\textcolor{black}{
\begin{equation}
\begin{array}{c}
\!\!\!\!\!\!\!\!\!\!\!\!\!\!\!\!\!\!\!\!\!\!\!\!\!\!\!\!\!\!\!\!\mathbf{RRM}:\underset{\xi_{q}^{\left(i,j\right)}\left(t\right)}{\max}\:\:\,\underset{q\in\mathcal{Q}}{\sum}\mathcal{N}_{q}\\
\textrm{s.t. \ensuremath{\left(\ref{eq:com}\right)}}-\left(\ref{eq:delay-1}\right).
\end{array}\vspace{-0.5em}
\end{equation}
}

\textcolor{black}{Due to the dynamic network environment, the proposed
RRM problem cannot be solved directly by traditional static optimization
tools. This paper takes advantage of DRL in combating dynamic network
environments and further combines the high adaptability requirements
}in practical deployment applications\textcolor{black}{{} to propose
the TF-DARM algorithm. The proposed algorithm will be introduced in
the next section.}\vspace{-0.5em}

\section{TF-DARM Algorithm Design\label{sec:TF-DARM-Algorithm-Design}\vspace{-0.5em}}

In this section, the TF-DARM algorithm for E2E service provisioning
in satellite networks is designed to solve the formulated RRM problem.
Specifically, we first analyze the service provisioning process and
model the RRM problem as the MDP. Subsequently, we present the training
and application of the TF-DARM algorithm. \vspace{-0.5em}

\subsection{\textcolor{black}{RRM Problem Conversion}\label{subsec:RMM-Problem-Conversion}\vspace{-0.5em}}

The RRM problem is essentially the process of achieving high-performance
E2E service provisioning by allocating the optimal available resources
for each service request. During the service provisioning, the P2P
transmission decision of a service request is based on the currently
available network resources and the distance between the next-hop
node and the destination node and will affect the next network environment,
which is an MDP. The main elements of the proposed MDP are shown as
follows:

\subsubsection{State--Network Resource and Service Orientation}

In the RRM problem, state evolution by time slot is a commonly adopted
way \cite{tsai2022multi}, and only one-hop transmission of a service
request is performed in each time slot, which is not in line with
the actual scenario. We design a three-dimensional state evolution
way and define the state $s_{q}^{i}\left(t\right)$ to characterize
the available resources of LEO and service orientation, as follows:\vspace{-0.3em}\textcolor{black}{
\begin{equation}
s_{q}^{i}\left(t\right)\!=\!\left[\overline{\mathcal{R}}^{\left(i,j\right)}\left(t\right)\!,\!\eta_{q}^{\text{\ensuremath{\left(i,j\right)}}}\!\left(t\right),\Delta_{q}^{j}\!\left(t\right),\!\chi_{q}^{j}\left(t\right)\right]_{j\in N_{i}}\!,\vspace{-0.5em}
\end{equation}
where $N_{i}$ is the set of next-hop nodes of LEO $i$, $\overline{\mathcal{R}}^{\left(i,j\right)}\left(t\right)$
is the normalized available communication resources of LEO $i$, and
$\eta_{q}^{\text{\ensuremath{\left(i,j\right)}}}\!\left(t\right)$
is the probability of successfully transmitting the service request
on ISL $\left(i,j\right)$. $\Delta_{q}^{j}\!\left(t\right)$ is the
service} orientation\textcolor{black}{{} information, i.e., the relative
position between the next-hop }node\textcolor{black}{{} $j$ and the
destination node $D_{q}$, including closing, far away, reaching,
etc., expressed as:}\vspace{-0.5em}\textcolor{black}{
\begin{equation}
\Delta_{q}^{j}\left(t\right)=\begin{cases}
2, & j=D_{q},\\
1, & j\neq D_{q}\textrm{ and }j\textrm{ is close to }D_{q},\\
-1, & j\neq D_{q}\textrm{ and }j\textrm{ is not close to }D_{q},\\
0, & \textrm{otherwise}.
\end{cases}\vspace{-0.5em}
\end{equation}
$\chi_{q}^{j}\left(t\right)$ is the ratio of supply and demand of
available communication resources of next-hop nodes, expressed as:}\vspace{-0.8em}

\textcolor{black}{
\begin{equation}
\chi_{q}^{j}\left(t\right)\!=\!\begin{cases}
1, & \!j=D_{q},\\
\!\frac{\underset{j'\in N_{j}}{\sum}\overline{\mathcal{R}}^{\left(j,j'\right)}\left(t\right)}{\left|\mathcal{Q}_{j}\left(t\right)\right|}, & \!j\!\neq\!D_{q}\textrm{\! and\! }j\in\widehat{N}_{i}\left(t\right),\\
0, & \!\textrm{otherwise}.
\end{cases}\vspace{-0.5em}
\end{equation}
where $\mathcal{Q}_{j}\left(t\right)$ is the set of service requests
in LEO $j$ in the $t$-th time slot, $\left|\cdot\right|$ }indicates\textcolor{black}{{}
getting the number of elements in a set, and $\widehat{N}_{i}\left(t\right)$
is the set of optional next-hop nodes.}

\subsubsection{Action--Regional Resource Management Strategy}

Based on the RRM mode, we design an action space $\mathcal{A}_{q}^{i}\left(t\right)$
that can take into account the fixed output dimensions of the neural
network and \textcolor{black}{the changing resource chains of the
dynamic network environments for E2E service provisioning, which corresponds
to the all next-hop nodes of LEO $i$ and can be generally expressed
as $\mathcal{A}_{q}^{i}\left(t\right)=N_{i}\cup None$,}\textcolor{blue}{{}
}where $None$ indicates that no next-hop node is selected, i.e.,
the service request is not transmitted. Furthermore, due to the intermittent
connectivity of ISLs, the set of available actions \textcolor{black}{$A_{q}^{i}\left(t\right)$
is variable and can be determined according to the connection relationship
of ISLs, i.e., $\!A_{q}^{i}\!\left(t\right)\!\!=\!\!\left\{ \!a_{q}^{i}\!\left(t\right)\!=\!j\left|j\!\in\!\widehat{N}_{i}\!\left(t\right)\!\right.\right\} \!\cup\!None$.}

\subsubsection{Reward--Phased Destination Guidance}

\textcolor{black}{Considering the P2P transmission decision-making,
the model of simply giving rewards upon reaching the destination node
makes effective guidance information too sparse. To this end, we design
the phased reward function matching the service orientation information
to progressively guide service requests to approach the destination
node, expressed as:}\vspace{-0.5em}\textcolor{black}{
\begin{equation}
r_{q}^{i}\left(t\right)=\begin{cases}
100 & \!\!a_{q}^{i}\left(t\right)\!=D_{q},\\
\!\frac{\Delta_{q}^{j}\left(t\right)\cdot\chi_{q}^{j}\left(t\right)}{L_{q}^{i}\left(t\right)}\!, & \!\!a_{q}^{i}\left(t\right)\!\neq\!D_{q},\!\textrm{\ensuremath{D_{q}}}\!\notin\!\widehat{N}_{i}\!\left(t\right)\!,\widehat{N}_{i}\!\left(t\right)\!\neq\!\emptyset,\\
0, & \!\textrm{otherwise}.
\end{cases}\vspace{-0.3em}
\end{equation}
}\vspace{-0.5em}

To sum up, the RRM problem can be converted into maximizing the long-term
cumulative reward, as follows:\vspace{-1.0em}

\begin{equation}
\max\,\mathbb{E}_{\pi}\left[\underset{t\in T}{\sum}\underset{q\in\mathcal{Q}}{\sum}\underset{i\in P_{q}\left(t\right)}{\sum}r_{q}^{i}\left(t\right)\right],\vspace{-0.5em}
\end{equation}
where $\pi$ represents a mapping from $s_{q}^{i}\left(t\right)$
to $a_{q}^{i}\left(t\right)$, i.e., $a_{q}^{i}\left(t\right)=\pi\left(s_{q}^{i}\left(t\right)\right)$.
$P_{q}\left(t\right)$ indicates the set of LEOs that transmit the
service request $q$ in the $t$-th time slot.\vspace{-0.5em}

\subsection{Training and Application of the TF-DARM Algorithm\label{subsec:Training-and-Application}\vspace{-0.3em}}

As we mentioned earlier, this paper expects to achieve high adaptability
of the algorithm in different network environments and avoid retraining.
Therefore, we choose to adopt a centralized training mode with the
NCC as the agent to obtain a trained model that is independent of
the network scale and introduce the Advantage Actor-Critic (A2C) framework
to optimize the parameters of the neural network. The overview of
training of the proposed algorithm is shown in Figure \ref{fig:An-overview-of}.

Specifically, the NCC observes the state of the local topology environment
centered on any LEO and applies the actor network $\pi_{\vartheta}\left(s_{q}^{i}\left(t\right)\right)$
to select the RRM strategy, where a mask mechanism is designed to
ensure the validity of the selected strategy. Then, the NCC collects
a series of experience data by interacting with the local topology
environments\textcolor{black}{. These }experience\textcolor{black}{{}
data }consist of\textcolor{black}{{} the three-dimensional state evolution
sequences, and provide rich local topological environment features,
which can help the NCC quickly and comprehensively learn the changing
network environments. Sample the }experience\textcolor{black}{{} data
and apply loss functions to update the parameters of the actor network
and critic network, where the critic network, denoted by }$V_{\varpi}\left(s_{q}^{i}\left(t\right)\right)$,
outputs the predicted reward and is responsible for evaluating the
selected strategy. \textcolor{black}{The loss functions are expressed
as follows:}\vspace{-0.5em}\textcolor{black}{
\begin{equation}
\mathcal{L}\left(\vartheta\right)=\!-\frac{1}{\left|\mathcal{M}\right|}\underset{s_{q}^{i}\left(t\right)\in\mathcal{M}}{\sum}\!\!\log\!\left(\pi_{\vartheta}\!\left(a_{q}^{i}\left(t\right)\left|s_{q}^{i}\left(t\right)\right.\right)\right)\cdot\mathcal{W}_{q}^{i}\left(t\right),\vspace{-0.5em}\label{eq:A_loss}
\end{equation}
}\vspace{-0.5em}\textcolor{black}{
\begin{equation}
\mathcal{L}\left(\varpi\right)=\frac{1}{2\cdot\left|\mathcal{M}\right|}\underset{s_{q}^{i}\left(t\right)\in\mathcal{M}}{\sum}\left(R_{q}^{i}\left(t\right)-V_{\varpi}\left(s_{q}^{i}\left(t\right)\right)\right)^{2},\vspace{-0.5em}\label{eq:C_loss}
\end{equation}
where $\mathcal{M}=\left\langle s_{q}^{i}\left(t\right),a_{q}^{i}\left(t\right),r_{q}^{i}\left(t\right),s_{q}^{i'}\left(t\right),d_{q}^{i}\left(t\right)\right\rangle $
is the minibatch, and $d_{q}^{i}\left(t\right)$ indicates whether
the service request $q$ continues to be served in the $t$-th time
slot. $\mathcal{W}_{q}^{i}\left(t\right)$ is the Temporal-Difference
error. $R_{q}^{i}\left(t\right)=r_{q}^{i}\left(t\right)+\gamma\cdot V_{\varpi^{-\!\!}}\left(s_{q}^{i'}\left(t\right)\right)$
is the estimated state value, and $\gamma$ is the discount factor.}

In the application phase, as shown in Figure \ref{fig:An-overview-of},
the trained model is deployed to each LEO of the satellite network
to make distributed decision-making about RRM strategies.\vspace{-0.3em}
\begin{figure}
\begin{centering}
\includegraphics[scale=0.25]{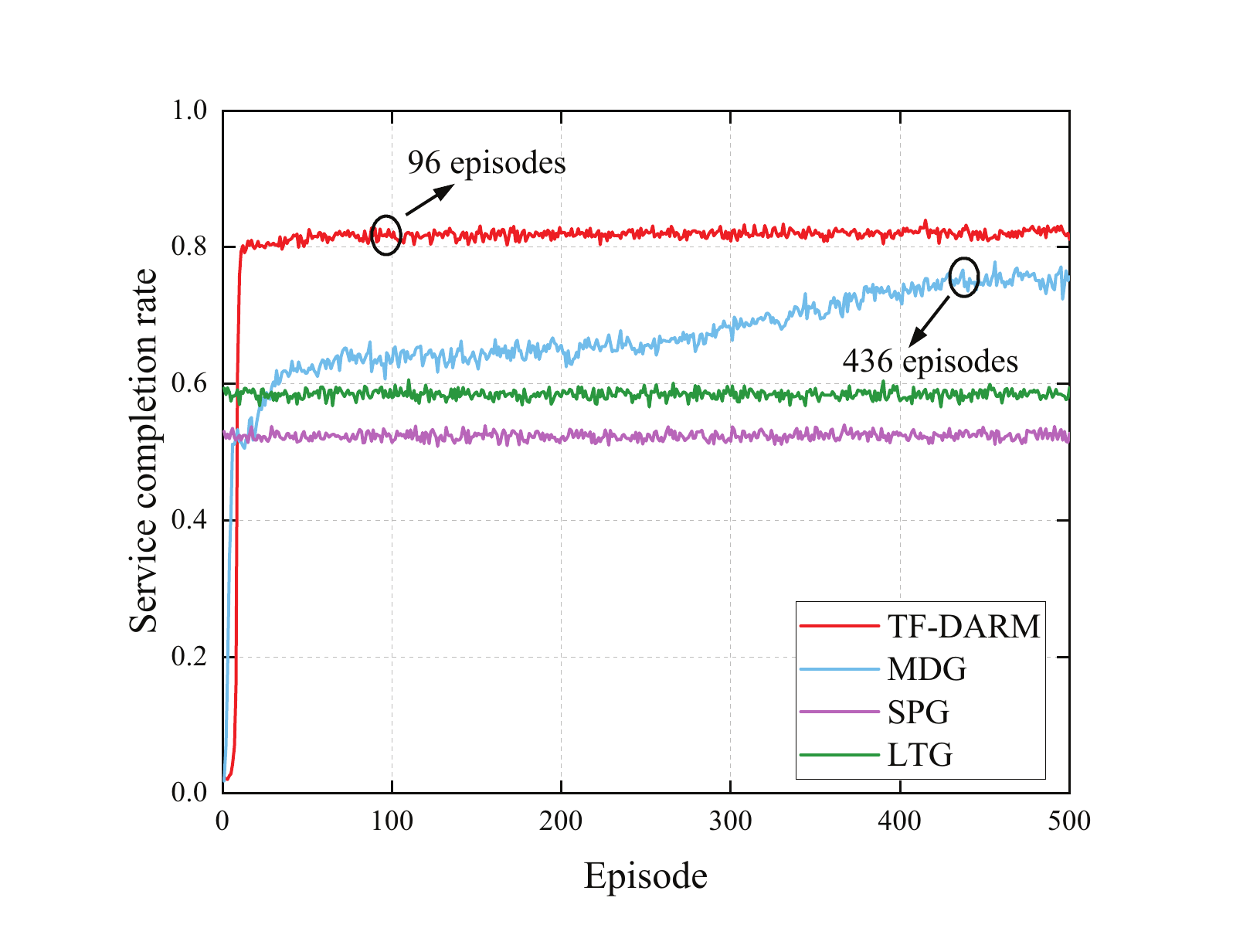}\vspace{-2.0em}
\par\end{centering}
\caption{\label{fig:Convergence-performance-of}Convergence performance of
the TF-DARM algorithm and comparison algorithms.}
\vspace{-1.5em}
\end{figure}

\section{Simulations\label{sec:Simulations}\vspace{-0.5em}}

In this section, we give the convergence result and the numerical
results for evaluating the performance of the TF-DARM algorithm. For
the simulations, the model is trained in a satellite network scenario
with 66 LEOs and tested in different satellite networks. Specifically,
according to \textcolor{black}{the Iridium communication system, }66
LEOs are distributed over six orbits at a height of 780 km and with
an inclination \textcolor{black}{of 86.4$^{\circ}$}. The configurations
of other test networks are set according to the Starlink constellation.
\textcolor{black}{The duration of the planning cycle is 1 hour from
30 Dec. 2024 10:00:00 to 30 Dec. 2024 11:00:00.} \textcolor{black}{For
transmission rate, we set $\mathcal{R}^{\left(i,j\right)}\left(t\right)\in\left[5,10\right]$Gbps.
Moreover, we set $\gamma=0.99$, $episode=500$, $\left|\mathcal{M}\right|=64$,
$\tau=60$s. The learning rates of actor and critic networks are set
$\alpha_{\vartheta}=2e^{-4}$ and $\alpha_{\varpi}=5e^{-4}$, respectively.
Besides, we set that service requests to arrive randomly, with more
than 100 arriving for each LEO, $B_{q}^{\textrm{req}}=5$Gbits and
$L_{q}^{\textrm{req}}=5$s.}\textcolor{blue}{} To compare the performance,
three additional approaches are considered: \vspace{-0.2em}
\begin{itemize}
\item \textbf{Mismatched Destination Guidance (MDG)}: This approach adopts
the framework of the proposed TF-DARM algorithm but does not provide
service orientation information that matches the phased reward function
of destination guidance, which may lead to long-term exploration and
unclear destinations.\vspace{-0.1em}
\item \textbf{Shortest Path Greedy (SPG)}: This approach selects the next-hop
node closest to the destination node when making a P2P transmission
decision for each service request.\vspace{-0.1em}
\item \textbf{Least Time Greedy (LTG)}: This approach selects a strategy
with the least single-hop delay when making a P2P transmission decision
for each service request.\vspace{-0.2em}
\end{itemize}
\setlength{\parindent}{1em}

We first evaluate the convergence performance of the TF-DARM algorithm,
as shown in Figure \ref{fig:Convergence-performance-of}. It can be
seen that at the beginning of the iteration, the TF-DARM and MDG algorithms
have a low service completion rate. With the number of iterations
increasing, the agent quickly learns effective decision-making information
and achieves a higher service completion rate than the SPG and LTG
algorithms. Besides, the TF-DARM algorithm can provide targeted destination
guidance to obtain better convergence performance and a faster convergence
rate than the MDG algorithm.

\begin{figure}
\begin{centering}
\includegraphics[scale=0.25]{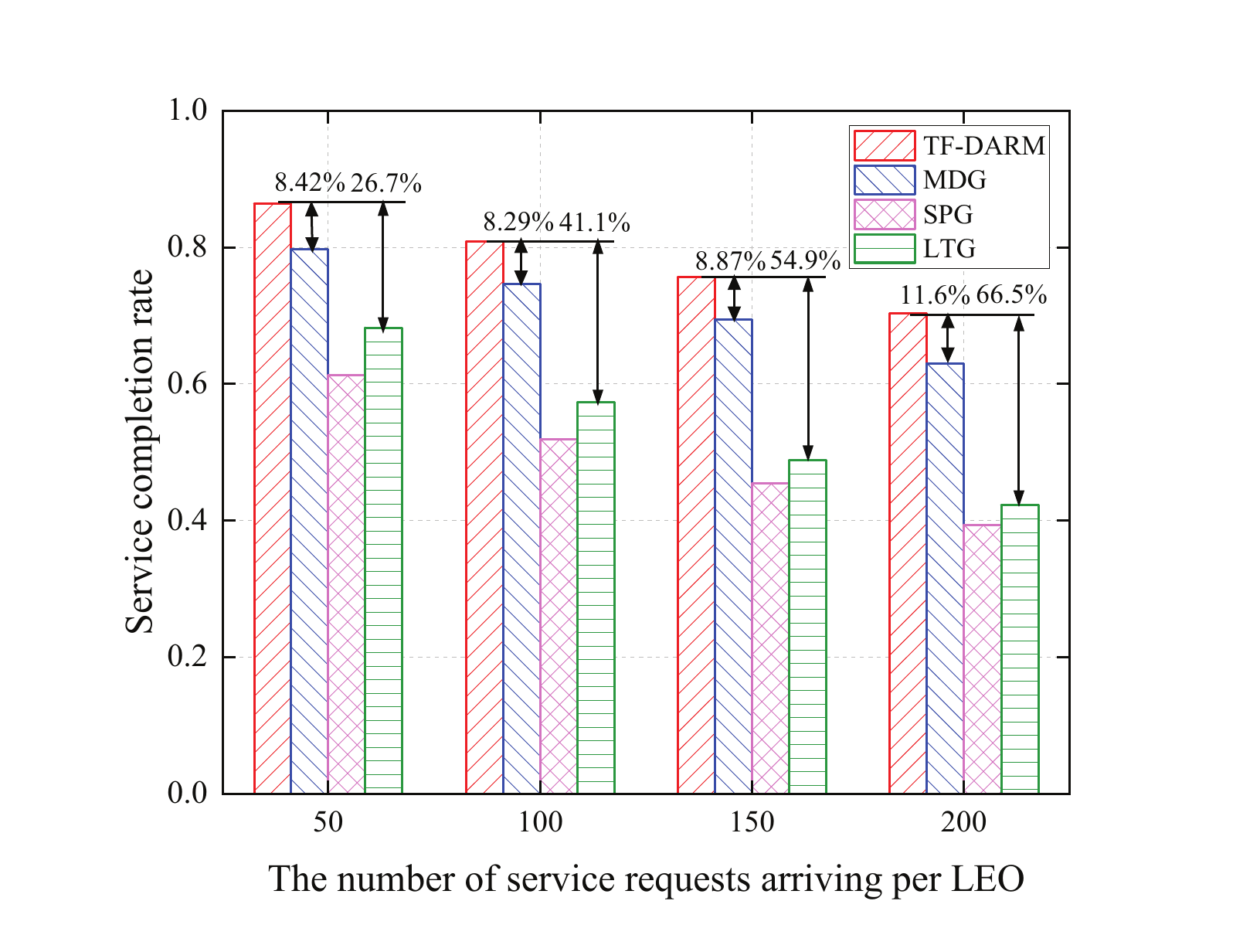}\vspace{-2.0em}
\par\end{centering}
\caption{\label{fig:Service-completion-rate}Service completion rate versus
different number of service requests.}
\vspace{-1.5em}
\end{figure}

Figure \ref{fig:Service-completion-rate} shows the service completion
rate under different number of service requests. As we expected, the
TF-DARM algorithm achieves the best service performance. Due to a
lack of clear service guidance and visionary strategy selection, comparison
algorithms have poor service effectiveness and low service completion
rates. When the number of service requests arriving per LEO is 50,
\textcolor{black}{the service completion rate of the TF-DARM algorithm
is higher than 8.42\% and 26.7\% compared with the MDG algorithm and
LTG algorithm, respectively. As the number of service requests increases,
the improvement is becoming more and more significant.} Besides, since
network resources are limited, as the number of service requests increases,
the service completion rate of all algorithms gradually decreases.

Finally, we test the service performance of the proposed algorithm
under different network scales, as shown in Table \ref{tab:Service-completion-rate}.
To ensure the fairness of the test, we set that the number of service
requests increases proportionally with the available resources of
different network scales. The results indicate that the proposed TF-DARM
algorithm exhibits better service performance when generalized to
satellite networks with varying numbers of LEOs. Compared with the
MDG algorithm, the SPG algorithm, and the LTG algorithm, \textcolor{black}{the
TF-DARM algorithm obtains the minimum gains of 2.7\%, 11.9\%, and
10.2\%, respectively.}\textcolor{blue}{{} }Furthermore, due to the orbital
deployment of LEOs, when the number of orbits is small, the ISLs of
two LEOs in different orbital planes may rarely be connected, which
causes a large number of timeouts for service requests with delay
requirements, resulting in poor service performance. With network
scale increases, the ISL between the two LEOs has better connectivity.
In this case, the service performance of the SPG algorithm has been
significantly improved and surpassed the LTG algorithm.\vspace{-0.3em}
\begin{table}
\caption{\label{tab:Service-completion-rate}Service completion rate on different
network scales.}
\vspace{-1.0em}
\centering{}%
\begin{tabular}{|c|c|c|c|c|c|}
\hline 
Network scale & 172 & 348 & 720 & 1584 & \multirow{2}{*}{\textcolor{black}{Minimum gain}}\tabularnewline
\cline{1-5} \cline{2-5} \cline{3-5} \cline{4-5} \cline{5-5} 
Configuration & 4{*}43 & 6{*}58 & 36{*}20 & 72{*}22 & \tabularnewline
\hline 
\hline 
TF-DARM & \textbf{\footnotesize{}\cellcolor{gray!25}}0.378 & \textbf{\footnotesize{}\cellcolor{gray!25}}0.569 & 0.837 & 0.834 & \textcolor{black}{/}\tabularnewline
\hline 
MDG & \textbf{\footnotesize{}\cellcolor{gray!25}}0.351 & 0.541 & 0.718 & 0.722 & \textcolor{black}{0.027}\tabularnewline
\hline 
SPG & 0.253 & \textbf{\footnotesize{}\cellcolor{gray!25}}0.450 & 0.616 & 0.618 & \textcolor{black}{0.119}\tabularnewline
\hline 
LTG & \textbf{\footnotesize{}\cellcolor{gray!25}}0.276 & 0.461 & 0.585 & 0.565 & \textcolor{black}{0.102}\tabularnewline
\hline 
\end{tabular}\vspace{-1.6em}
\end{table}

\section{Conclusion\label{sec:Conclusion}\vspace{-0.3em}}

In this paper, we explore the topology features of satellite networks
and adopt the designed RRM mode to formulate the RRM problem for E2E
service provisioning to obtain the unified decision space. Subsequently,
we model the service provisioning process as the MD\textcolor{black}{P,
and based on the A2C framework, we propose the TF-DARM algorithm to
combat the dynamic network environments and avoid retraining. }The
proposed algorithm adopts the three-dimensional state evolution way
and leveraging designed generalized action space, it can take into
account the fixed output dimension of the neural network and the changing
resource chains for E2E service provisioning. Furthermore, the matched
design of the service orientation information and phased reward function
effectively improves the service performance of the algorithm. \textcolor{black}{Simulation
results demonstrate that the TF-DARM algorithm has the best convergence
performance and fastest convergence rate and achieves highly adaptive
resource management for varying network scales to boost practical
deployment applications. }\vspace{-0.6em}

\bibliographystyle{IEEEtran}
\bibliography{reference}

\end{document}